\baselineskip .47cm
\magnification 1200

\vfill
\centerline{\bf Democratic Reinforcement: Learning}
\centerline{\bf via Self-Organization}
\bigskip
\centerline{Dimitris Stassinopoulos{$\ast$} and Per Bak{$\dag$}}
\medskip
\centerline{\it {$\ast$}Center for Complex Systems, Florida Atlantic
University}
\centerline{\it Boca Raton, Florida 33431, USA}
\centerline{\it {$\dag$}Department of Physics, Brookhaven National Laboratory}
\centerline{\it Upton, New York 11973, USA}
\bigskip
\bigskip
\noindent{\it ABSTRACT}\smallskip
{\bf
The problem of learning in the absence of external intelligence is
discussed in the context of a simple model.
The model consists of a set of randomly connected, or layered integrate-and
fire neurons. Inputs to and outputs
from the environment are connected randomly to subsets of neurons.
The connections between firing neurons are strengthened or weakened
according to whether the action is successful or not.
The model departs from the traditional gradient-descent based approaches
to learning by operating at a highly susceptible ``critical''
state, with low activity and sparse connections between firing neurons.
Quantitative studies on the performance of our model in a simple
association task show that by tuning our system close to this critical
state we can obtain dramatic gains in performance.
}
\bigskip
\bigskip
\noindent{\bf I. Introduction}\smallskip

One of the most remarkable properties of biological neural networks
is their ability to learn via {\it self-organization}.
Simply put, this means that animals acquire experience and make sense of
their environment without the aid of a ``teacher'' or some other form of
external intelligence.
To any non-expert that has ever seen a toddler acquiring a new skill with
virtually no guidance or an animal adapting to a novel situation
the statement would seem obvious. Yet for all its simplicity and common sense
this idea has long remained on the fringe of the experimental and theoretical
research concerning brain function, in all likelihood because of the
severity of the contraints it imposes on brain modelling.

The term self-organization has been used in many different
disciplines such as physics, chemistry, biology, and psychology, and often
to convey different underlying mechanisms. Here, however, we will discuss
self-organization solely in the context of learning$^{1}$.

The oldest and perhaps still most dominant approach to understanding the brain
is what we call the ``engineered brain'' paradigm.
According to this paradigm, brain function emerges because nature,
in the role of the engineer,
has created all the necessary mechanisms by
establishing an intricate web that brings billions of
pieces together.
But how can evolution achieve such an engineering
feat? We do not deny the role of evolution in many aspects of brain
function -- the very fact that our brain is different than a lobster's
brain has to be attributed to evolution. Nevertheless, it cannot possibly
account for the brain's ability to deal with unforseen situations, specific
to an individual's experience, or for novel ones that
evolution had never had the opportunity to confront.

In providing an alternative to this view,
the field of artificial neural networks (ANN) has been instrumental
(For reviews see Hertz {\it et
a}l$^2$ and Haykin$^3$). The major
contribution of ANN was that it demonstrated how non-trivial
tasks can be achieved with networks composed of many simple computing
elements. It also offered the first evidence that principles for brain
function can be captured with models that have simple structure.
Despite all the important insights ANN offered, however, they have not
eliminated the need for an external intelligence. In the widely used
{\it supervised learning} paradigm this takes the form of a ``teacher''
providing the system with
a detailed scheme for the update of the synaptic weights
based on knowledge of the goal to be achieved.  Furthermore, most
models for learning use gradient-based update rules, such as
back-propagation, which are biologically implausible because they impose
strong constraints on the architecture and they require computation that
cannot be performed by the neural network itself. Thus, again the network
is formed by design rather than by self-organization.

The issues of self-organization have been addressed in the context of
reinforcement learning models$^{2,3}$. These models are more realistic
in the sense that there is no teacher explaining how to modify the
synaptic weights, but only a ``critic'' telling the system whether its
performance is successful or not. Most reinforcement learning models,
however, still rely on back-propagation$^4$, or some other overseeing agency
possessing prior knowledge of the problem, for the update of the synaptic
weights.
There is one exception, however. Barto$^{5}$, in one of the first variants
of his {\it Associative Reward-Penalty}$^6$ ($A_{R-P}$) algorithm, discusses
the idea of ``self-interested'' elements which do not have access to
information other than a feedback signal from the environment
broadcast simultaneously to all elements.
We very much agree with his
view that the difficulties in solving the problem of learning
under the severe constraints imposed by self-organization are fundamental.

Recently, we have proposed {\it Democratic Reinforcement} ($DR$) as a
new approach to the long-standing issues of learning via
self-organization$^{7}$. A similar approach was originally used to
solve a non-trivial tracking problem by a continuous modification of
its synaptic weights$^{8}$.
An evaluative feedback is sent democratically to all neurons simultaneously.
The reinforcement rule operates in two modes: a ``learning''
mode when the evaluative feedback is positive, and an
``exploration'' mode when the evaluative feedback is negative.
The reinforcement rule depends on the firing states of
the presynaptic and postsynaptic neurons only. The novel
feature of our model
is that the threshold for firing is
regulated in order to keep the output activity minimal. This sets the system
up at or near a critical state, which turns out to be crucial for the
performance of the network.

To the best of our knowledge, the $A_{R-P}$ and
 the $DR$ represent the only attempts
to address the problem of learning via self-organization.
However, the two algorithms are fundamentally
different. While the $A_{R-P}$ is a gradient-descent algorithm,
$DR$ solves problems by operating at or near a highly-susceptible ``critical''
state in which the system becomes very sensitive to modification of the
synaptic weights.
To make our point more concrete we shall discuss the two algorithms in the
context of a specific association task.
\bigskip
\bigskip
\noindent{\bf II. The Model}\smallskip

The problem we are addressing can be summarized as follows:
How can many ``agents'', be it neurons or some other kind of computing
element, operating under local rules and receiving input only from a small
fraction of other agents, cooperatively perform macroscopic tasks imposed
on them by their environment?
For the agents to perform a task, they are given the freedom to tune
some local parameters,
such as their synaptic weights, to appropriate values.
The severity of the problem arises from the requirements: i) that
there can be no external intelligence with prior knowledge of the problem to
instruct the agents on how to tune their parameters, only some
form of overall evaluative feedback that tells the agents when some
parameters essential for ``survival'' fall out of bounds, and
ii) that the system is robust and versatile, allowing for solutions that
are not task or architecture-specific.
 From an agent's point of view any effective adjustment can be made only when
the adjustment has a detectable impact on the overall, collective behavior
of the system, allowing for a robust feedback. This, however,
 is a non-trivial requirement. Most neural networks
are rather insensitive to small changes in their
parameters.
Then we are left with a situation where no agent can
``learn'' from its actions because there is no way for it to know
whether it should get ``credit'' or ``blame'' in the final evaluation.
This is known as the {\it credit-assignment problem}{$^{9}$}.

The above problem remains largely unresolved.
Of course, real biological
networks serve as examples par-excellence that solutions to this problem
exist, but how do we mimic this in a simple model?

Our model imposes no constraints on the architecture whatsoever.
In the most general case neurons are connected randomly to each
other via unidirectional connections$^{7}$. Each neural unit receives
input and
sends output to a small fraction of the total number of units. This ensures
that the majority of units in our system are hidden,
i.e., units that interact with the environment indirectly through other units.
The system interacts with the outer world in three ways (Fig. 1);
via i) its input units which receive input from the outside and thus
provide the system with information about the state of the world;
ii) its output units which allow the system to act on the environment;
and iii) a binary yes/no feedback that is broadcast
to all units and indicates whether the action to the environment
was successful or not.

Although from a conceptual point of view the random architecture is
the most appealing,
we found the layered architecture of Fig.1 to work better$^{10}$.
Here each unit is connected to its three nearest neighbors in
the next layer. The units can be either in a firing state, $n_i=1$, or in
a quiescent state, $n_i=0$. In standard fashion, each unit integrates the
input of its presynaptic units, $h_i=\sum_j J_{ij}n_j$. The unit fires
if the input exceeds a threshold $T$. Input patterns, indicated in
Fig. 1 by dark disks, are presented to the system by setting the
corresponding units into a firing state. The system acts on the environment
via its output sites, shaded disks in Fig. 1. The
feedback, $r$, broadcast by the environment, takes two values; positive, $r_0$,
if the action was evaluated as successful and negative, $-r_0$, if not.
Both the synaptic modification and the regulation of the threshold depend
on the evaluative feedback, $r$.

The update rule affects only connections between firing neurons,
$J_{ij} \to J_{ij}  + [rJ_{ij}+h_{ij} ] n_i' n_j$ , where
$n_i'$ denotes the state of the $i$'th neuron at the next time step and
$h_{ij}$ is a random noise between $-h_0$ and $h_0$. The
outgoing weights are normalized,
$J_{ij} \to J_{ij} / \sum_iJ_{ij}$. The rule differs from standard
gradient-descent based update rules in one crucial aspect: When $r>0$ the
system operates in a ``learning'' mode in which connections are being
strengthened and the performance improves but when $r<0$ the system
operates in an ``exploratory'' mode in which strong connections
are being weakened and weaker connections are being strengthened. Typically,
during this phase the performance deteriorates. In contrast, standard
reinforcement schemes, such as $A_{R-P}$, rely on an improvement of
the performance both for positive and negative $r$
and perform the exploration stochastically.

In addition to the synaptic update rules, $r$ regulates the threshold, $T$.
The objective is that the output activity is kept to a minimum.
This is essential to ensure that the system attributes credit and blame to
the minimum possible number of active units, in order to keep the network
intact for other problems. In our first versions of the $DR$
algorithm (Ref. 7, 8) the output activity was regulated to a small but
arbitrarily chosen level of activity. Choosing a value for this parameter,
however, assumes prior knowledge of the task to be completed which runs
counter to the self-organization philosophy. More recently$^{11}$ we have
introduced a threshold mechanism that depends solely on $r$,
$T \to T+\delta (r)$, where $\delta$ assumes a positive value, $\delta_+$,
if $r>0$, and a negative value, $\delta_-$, if $r<0$ ($|\delta_-|>>\delta_+$).

The typical criterion for success, $r>0$,
is that the selected output sites are active and for failure,
$r<0$, that
at least one of the selected output sites is inactive.
On first thought, that might sound like nonsense:
the system can trivially obtain a positive feedback and thus get its
reward by lowering its threshold, for instance, and keeping all of
its output units active! However, the solution that the system opts for is
the one where the selected output sites are active and all the rest
are inactive.
But even if we accept that such a rule makes computational sense
what sense does it make in terms of biology? It is true that in some
simple situations we may view this reward/penalty coming from the
environment. This was the case in Ref. 7 where we used the analogy
of a monkey that presses one or more buttons. In such a situation
the environment indeed provides food as long as the selected buttons are
among the ones pressed by the monkey. However, the monkey can not be considered
successful merely because now and then it happens to press the right buttons.
It
is important to reduce the incorrect actions.
In that respect it makes more sense to view $r>0$
not as an external reward but rather as the default mode of operation;
an innate tendency of the system to minimize its efforts, while still having
success.
In contrast, one might view the $r<0$ signal as an external wake-up call
announcing that
something is wrong, for instance when some parameter that is
crucial to survival exceeds a certain value.
The preference for passivity
is sharply interrupted when $r<0$. There is no symmetry between $r<0$ and
$r>0$.

How does the system solve problems? How does it
successfully attribute credit and blame where it is due?
Our studies indicate that this involves a build-up process in which
the synaptic ``landscape'' reaches a near critical, highly susceptible
state in which small changes in the synaptic weights can have a big effect
on the collective activity. In such a state the
system can establish efficiently causal relationships between
changes in the synapses and the output.
To achieve such a critical state the system: i) assigns credit and blame
only to connections between active neurons, ii) keeps the activity low by
means of the global regulation of the threshold and the local learning
rules. By combination of these two mechanisms the system
 attributes credit and blame selectively by driving
the system to the interface between success and failure.
\bigskip
\bigskip
\noindent{\bf III. Self-organization in a Simple Association Task}\smallskip

In an association task we ask the system to generate a certain input/output
pattern.
The insets in Figs. 2a, b offer examples of a simple association task.
The system accomplishes the task by ``carving'' paths between the input
sites and the output sites. In previous work$^{7,8,12}$ we have
investigated the performance of $DR$ in a variety of situations: multiple
input/output patterns, recovery from ``damage'', tracking, conditioning,
and so on.
Here we will be concerned with the question of degradation of
performance as the size of the association task
grows.

We consider an $L_1\times L_2$ layered network (Fig. 2a, inset).
The number of layers in this networks is kept fixed, $L_1=16$, while
the lateral dimension is varied, $L_2=16, 32, 64,\dots$ The number, $c$,
of input and output sites in the input/output pattern is varied
accordingly, $c=L_2/16$. Here the input is confined to the top
row and the output to the bottom row.
To minimize crossover between paths
we keep the input and output sites in pairs, well separated from each
other. More precisely, each of the $c$ columns of a given network
contains a single input and a single desired output site.

One motivation for this analysis is to demonstrate in a convincing
manner the difference between $DR$ and $A_{R-P}$. We were not so interested
in the absolute performances of the two algorithms, which tend to be sensitive
to the tuning of the various parameters, but rather to the scaling of the
performance with the size of the network.
For our simulations with $A_{R-P}$ the same layered architecture was chosen
(Fig. 1) but the number of layers was set to $L_1=4$
(more layers would degrade the performance of $A_{R-P}$ too much).
The lateral dimension is varied, $L_2=4, 8, 12,\dots$ The input/output
patterns were chosen with similar considerations as in the $DR$ case
(see insets in Figs. 2a,b for a comparison).
Figures 2a and 2b show examples of the performance, $P$, for $DR$ and
$A_{R-P}$ as a function of time. $P$ is defined as the temporal average of
the activity at the selected output sites minus the activity at the rest
of the output sites. Appropriate normalization assures that best performance
($P=1$) corresponds to persistent firing at the selected output sites only,
whereas worst performance ($P=-1$) corresponds to persistent firing everywhere
except at the selected sites.

The $DR$ is characterized by intervals of
rapid improvement in performance, interrupted by sudden dips. This behaviour
is a signature of the dual mode of operation of the algorithm: i) the
learning or exploitation mode in which the system strengthens connections
and ``weeds out'' irrelevant paths, and ii) the exploration mode in which the
system tends to spread the activity in an attempt to explore new possibilities
with subsequent decrease in the performance $P$. The $A_{R-P}$ performance
versus time, although also highly irregular, seems to have a very different
structure. It is dominated by very fast fluctuations
at the smallest time scale (not seen here due to averaging of $P$).
At longer time scales it seems to be dominated by long periods during
which the system seems trapped at a certain level of performance.
Once the system escapes this barrier the transition to a new performance
level appears to be very fast. We would like to point out that at the
individual
level, and with the limited information available to it, each neuron
{\it always} opts for the change that it expects will increase the collective
performance. In its decisions, however, it cannot take into account the
positive or negative contributions of the other neurons. Therefore, it is
only in a statistical sense that the system senses the gradient towards a
better performance and can tune its synapses accordingly.
The stochastic nature of the $A_{R-P}$ can also be witnessed in Fig. 3,
($\triangle$). Here we depict the time to completion of the task,
$\bar{t_s}$ (averaged over many runs obtained with different initialization)
as a function of the number, $c$, of input/output pairs.
In a first order approximation, it seems that $\bar{t_s}$ scales exponentially
with $c$, $\bar{t_s}\sim e^{\alpha c}$, with $\alpha\simeq 1.6$.

$DR$ (Fig. 3, $\bigcirc$) has a significantly better scaling behavior.
When plotted in a log-log plot (inset of Fig. 3)
$\bar{t_s}$ might follow a power law, $\bar{t_s}\sim c^{\gamma}$,
which subsequently breaks down around $c=8$.
If this is true it would not be inconsistent with our suggestion that the
algorithm operates
near a ``critical'', highly susceptible regime. Although evidence of such
a critical regime have previously been seen in the dynamics of our
system$^{7}$, this scaling behavior offers the first quantitative
evidence. Clearly, this initial data seems to be amenable to more than
one interpretation, therefore it is imperative that the direct consequences
of our critical-state hypothesis are tested further.

The convergence towards the critical state is accomplished by ensuring that
the patterns of activity for different input signals do not overlap, on one
hand, while, on the other hand, not being too sparse to connect inputs with
desired outputs. In ``sand'' models of self-organized criticality,$^{13}$
overlap of events (``avalanches'') is avoided by keeping the input rate low.
Here, criticality is achieved by keeping the output low.

It turns out that one can improve the efficiency, and carry the
system closer to the critical point by further ensuring that changes
in the activity, due to threshold modulation do not overlap in time, while
not happening too infrequently with respect to the synaptic modification rate.
We do so by allowing a variable rate
$\delta_+$ for increasing the threshold $T$. More precisely the rise of the
threshold is governed by $\delta_+(t,r)$, $\delta_+\to a\delta_+$,
where $a>1$. Notice that now $\delta_+$ is time dependent
since, while $r>0$, it is constantly increased and $r$ dependent
since it is reset to a small value whenever $r$ becomes negative,
$\delta_+\to a^{-L_1}\delta_+$. The rate of decrease of $T$, $\delta_-$ is
kept constant as before. The modified algorithm leads to a significant
improvement of the performance (Fig. 3, ($\sqcap$)).
Furthermore, the new curve, $\bar{t_s}(c)$, gives a stronger indication for
the existence of a power law with exponent $\gamma\simeq 1.3$.
The modification was chosen for its simplicity rather than its
performance and, based on our experience,
it appears to be straightforward to obtain further improvements.
\bigskip
\bigskip
\noindent{\bf IV. Conclusions}\smallskip

In this paper we have been concerned with the issues of self-organization
which must play a central role in brain function.
We propose a new mechanism through which efficient self-organized learning
takes place. The central element of is a build-up process that
allows the system to operate at a ``critical'' state, characterized
by high sensitivity to small modifications of the synaptic weights and low
output activity. The combination of those features allows the system to
establish strong cause-effect relationships that allow
the coexistence of many input/output patterns.

We suggest that the mechanisms that
enables self-organizion in our model might also underlie real brain function,
so $DR$ can serve as an excellent testbed for further exploration of the
consequences of such a hypothesis. It might be worthwhile exploring
whether some signature of the critical state described in our model
can also be observed in actual experiments.

\vfill
\eject
{\bf References}\smallskip

\noindent\item{1.} A qualification is necessary here to avoid
confusion of our use of the term ``self-organization'' as this used in
{\it unsupervised learning}. In this paradigm the system modifies
its synaptic ways with no feedback or any supervision whatsoever
from the environment. Our view is that the consideration of
environmental constraints is essential in biological learning and intelligent
behavior in general, therefore, unsupervised learning has been
entirely omitted from our discussion.
\medskip
\noindent\item{2.} J. Hertz, A. Krogh, \& R. G. Palmer, {\it Introduction
to the Theory of Neural Computation} (Addison-Wesley, Redwood, 1991).
\medskip
\noindent\item{3.} S. Haykin, {\it Neural Networks, a Comprehensive Foundation}
(Macmillan, New York, 1994).
\medskip
\noindent\item{4.} Based on the partial information
provided by the critic a target pattern is determined and the output-weight
errors computed.
The rest of the weights can then be updated by back-propagating this
error-signal through the network (see Hertz {\it et al} in Ref.~2).
\medskip
\noindent\item{5.}A. G. Barto, Human Neurobiology {\bf 4}, 229 (1985).
\medskip
\noindent\item{6.} A. G. Barto \& P. Anandan, IEEE Transactions on Systems,
Man, and Cybernetics, {\bf 15}, 360 (1985).
\medskip
\noindent\item{7.} D. Stassinopoulos \& P. Bak, Phys. Rev. E {\bf 51},
5033 (1995).
\medskip
\noindent\item{8.} P. Alstr\o m \& D. Stassinopoulos, Phys. Rev. E {\bf 51}
5027 (1995).
\medskip
\noindent\item{9.} M. L. Minski, {\it Proceedings of the Institute of
Radio Engineers} {\bf 48}:8-30.
\medskip
\noindent\item{10.} This is primarily due to computer time limitations
since the layered architecture gives substantially better
performance$^{7}$ but also due to our interest in making
comparisons with the $A_{R-P}$ algorithm which is implemented
in feedforaward architectures.
\medskip
\noindent\item{11.} D. Stassinopoulos \& P. Bak, for submission to
J. Theoret. Biol.
\medskip
\noindent\item{12.} D. Stassinopoulos, P. Bak, \& P. Alstr\o m,
in {\it Proceedings of the Second Appalachian Conference on Behavioral
Neurodynamics - Radford}, edited by K. Pribram, (Lawrence Erlbaum, New Jersey,
1994), 172-195.
\medskip
\noindent\item{13.} P. Bak, C. Tang, \& K. Wiesenfeld, Phys. Rev. Lett.
{\bf 59} 381 (1987).
\vfill
\eject
{\bf Figure captions}\smallskip

\item {Figure 1.} Block diagram of the model, here shown for the
layered architecture. The input sites that receive signals from
the environment are shown as dark discs. The output sites are
shown as shaded disks. Periodic boundary conditions are
assumed for the layers.
\medskip
\item {Figure 2.} a) $DR$: Performance vs. time,
for a $16\times 64$ system and, for the input/output pattern shown
in the inset.  Light sites denote quiescent units and dark sites denote
firing ones.
The firing sites connect the input and output sites by effectively forming
 wires. The parameters of the algorithm have been set to:
$r_0=0.1,\delta_+=0.01/16,\delta_-=-0.05/16$, and $h_0=0.01$.
The performance is obtained by averaging over $50$ time steps.
b) $A_{R-P}$: Performance vs. time (measured in `trials'), for a
$4\times 16$ system, and for the input/output pattern shown in the inset.
The connections between input and output sites are more complicated.
The central element of the algorithm is the update rule for the synaptic
weights, $J_{ij}\to J_{ij}+\eta(r[n_i-<n_i>]+\lambda(1-r)[-n_i-<n_i>])n_j$,
where $\eta$ is the `learning' coefficient, $\lambda$ is the `penalty
learning rate factor', $<n_i>=\tanh(\beta\sum_jJ_{ij}n_j)$ is the average
firing state, and $r$ is the feedback
(for details see Hertz {\it et al} in Ref. 2).
The parameters have been set to: $\eta=0.5, \lambda=0.001$, and $\beta=0.5$.
The performance is obtained by averaging over $100$ trials.
{\it Insets:} Typical successful ($P=1$) activity patterns for $DR$ and
$A_{R-P}$.
\medskip
\item {Figure 3.} Average time elapsed, $\bar{t_s}$, to completion of
an association task vs. number of input/output pairs, $c$. ($\triangle$)
$A_{R-P}$: systems of size $L_1\times L_2$,
$L_1=4$ and $L_2=4, 8, 12,$ and $16$ have been considered. For each case twenty
runs were performed (with the exception of the $4\times16$ system for which
we conducted five runs only, due to computing time limitations) for the same
association task but with different initialization; ($\bigcirc$) $DR$:
systems of size $L_1\times L_2$,
$L_1=16$ and $L_2=16, 32, 64, 128,$ and $192$ have been considered. For the
same association task fifty runs, with differing initializations, were
considered; ($\sqcap$) $DR$ with variable $\delta_+$ ($a=1.05$): systems of
size $L_1\times L_2$, $L_1=16$ and $L_2=16, 32, 64, 128, $ and $256$ have been
considered. For the same association task, fifty runs with differing
initializations, were considered. Vertical bars denote standard deviation.
{\it Inset:} Same data in log-log plot. The solid lines represent
least-squares fits. ($\triangle$): $\sim e^{1.6c}$; ($\sqcap$): $\sim c^{1.3}$.
\bye